\documentclass[12pt]{article}
\usepackage{amsfonts}
\usepackage{amssymb}
\usepackage{graphics,psboxit,amsmath}
\usepackage{subfigure}
\usepackage{graphicx}

\def\hybrid{\topmargin -30pt    \oddsidemargin 0pt 
        \headheight 0pt \headsep 0pt
        \textwidth 6.25in       
        \textheight 9.5in       
        \marginparwidth .875in
        \parskip 5pt plus 1pt   \jot = 1.5ex}

\hybrid

\def\baselinestretch{1.2}

\catcode`\@=11

\def\marginnote#1{}
\newcount\hour
\newcount\minute
\newtoks\amorpm
\hour=\time\divide\hour by60
\minute=\time{\multiply\hour by60 \global\advance\minute by-\hour}
\edef\standardtime{{\ifnum\hour<12 \global\amorpm={am}%
        \else\global\amorpm={pm}\advance\hour by-12 \fi
        \ifnum\hour=0 \hour=12 \fi
        \number\hour:\ifnum\minute<10 0\fi\number\minute\the\amorpm}}
\edef\militarytime{\number\hour:\ifnum\minute<10 0\fi\number\minute}

\def\draftlabel#1{{\@bsphack\if@filesw {\let\thepage\relax
   \xdef\@gtempa{\write\@auxout{\string
      \newlabel{#1}{{\@currentlabel}{\thepage}}}}}\@gtempa
   \if@nobreak \ifvmode\nobreak\fi\fi\fi\@esphack}
        \gdef\@eqnlabel{#1}}
\def\@eqnlabel{}
\def\@vacuum{}
\def\draftmarginnote#1{\marginpar{\raggedright\scriptsize\tt#1}}

\def\draft{\oddsidemargin -.5truein
        \def\@oddfoot{\sl preliminary draft \hfil
        \rm\thepage\hfil\sl\today\quad\militarytime}
        \let\@evenfoot\@oddfoot \overfullrule 3pt
        \let\label=\draftlabel
        \let\marginnote=\draftmarginnote
   \def\@eqnnum{(\theequation)\rlap{\kern\marginparsep\tt\@eqnlabel}%
\global\let\@eqnlabel\@vacuum}  }

\def\preprint{\twocolumn\sloppy\flushbottom\parindent 2em
        \leftmargini 2em\leftmarginv .5em\leftmarginvi .5em
        \oddsidemargin -.5in    \evensidemargin -.5in
        \columnsep .4in \footheight 0pt
        \textwidth 10.in        \topmargin  -.4in
        \headheight 12pt \topskip .4in
        \textheight 6.9in \footskip 0pt
        \def\@oddhead{\thepage\hfil\addtocounter{page}{1}\thepage}
        \let\@evenhead\@oddhead \def\@oddfoot{} \def\@evenfoot{} }

\def\numberbysection{\@addtoreset{equation}{section}
        \def\theequation{\thesection.\arabic{equation}}}

\def\underline#1{\relax\ifmmode\@@underline#1\else
        $\@@underline{\hbox{#1}}$\relax\fi}

\def\titlepage{\@restonecolfalse\if@twocolumn\@restonecoltrue\onecolumn
     \else \newpage \fi \thispagestyle{empty}\c@page\z@
        \def\thefootnote{\fnsymbol{footnote}} }

\def\endtitlepage{\if@restonecol\twocolumn \else \newpage \fi
        \def\thefootnote{\arabic{footnote}}
        \setcounter{footnote}{0}}  

\catcode`@=12
\relax

\def\figcap{\section*{Figure Captions\markboth
        {FIGURECAPTIONS}{FIGURECAPTIONS}}\list
        {Figure \arabic{enumi}:\hfill}{\settowidth\labelwidth{Figure
999:}
        \leftmargin\labelwidth
        \advance\leftmargin\labelsep\usecounter{enumi}}}
 \relax
\def\tablecap{\section*{Table Captions\markboth
        {TABLECAPTIONS}{TABLECAPTIONS}}\list
        {Table \arabic{enumi}:\hfill}{\settowidth\labelwidth{Table
999:}
        \leftmargin\labelwidth
        \advance\leftmargin\labelsep\usecounter{enumi}}}
 \relax
\def\reflist{\section*{References\markboth
        {REFLIST}{REFLIST}}\list
        {[\arabic{enumi}]\hfill}{\settowidth\labelwidth{[999]}
        \leftmargin\labelwidth
        \advance\leftmargin\labelsep\usecounter{enumi}}}
 \relax
\makeatletter
\newcounter{pubctr}
\def\publist{\@ifnextchar[{\@publist}{\@@publist}}
\def\@publist[#1]{\list
        {[\arabic{pubctr}]\hfill}{\settowidth\labelwidth{[999]}
        \leftmargin\labelwidth
        \advance\leftmargin\labelsep
        \@nmbrlisttrue\def\@listctr{pubctr}
        \setcounter{pubctr}{#1}\addtocounter{pubctr}{-1}}}
\def\@@publist{\list
        {[\arabic{pubctr}]\hfill}{\settowidth\labelwidth{[999]}
        \leftmargin\labelwidth
        \advance\leftmargin\labelsep
        \@nmbrlisttrue\def\@listctr{pubctr}}}
 \relax
\makeatother
\newskip\humongous \humongous=0pt plus 1000pt minus 1000pt

\newif\ifdtup

\relax

\def\be{\begin{equation}}
\def\ee{\end{equation}}
\def\ba{\begin{eqnarray}}
\def\ea{\end{eqnarray}}

\def\del{\partial}

\def\a{\alpha}

\def\g{\gamma}
\def\G{\Gamma}
\def\d{\delta}

\def\e{\epsilon}

\def\P{\Pi}

\def\th{\theta}
\def\Th{\Theta}
\def\m{\mu}
\def\n{\nu}
\def\om{\omega}
\def\Om{\Omega}
\def\l{\lambda}

\def\s{\sigma}
\def\S{\Sigma}

\def\cN{{\cal N}}

\def\no{\noindent}

\def\qq{\qquad}

\def\IR{\relax{\rm I\kern-.18em R}}

\def \ha {{1\over 2}}

\def \ov {\over}

\def\II{\relax{\rm 1\kern-.35em1}}
\def\IR{\relax{\rm I\kern-.18em R}}
\def\inv{^{\raise.15ex\hbox{${\scriptscriptstyle -}$}\kern-.05em 1}}

\def\tL{{\tilde L}}

\begin{document}
\renewcommand{\theequation}{\arabic{equation}}


\newcommand{\beq}{\begin{equation}}
\newcommand{\eeq}[1]{\label{#1}\end{equation}}
\newcommand{\ber}{\begin{eqnarray}}
\newcommand{\eer}[1]{\label{#1}\end{eqnarray}}
\newcommand{\eqn}[1]{(\ref{#1})}
\begin{titlepage}
\begin{center}

\hfill CERN-PH-TH/2005-230\\
\vskip -.1 cm
\hfill hep--th/0512158\\

\vskip .5in

{\Large \bf On supersymmetry and other properties of a class \\
of marginally deformed backgrounds}

\vskip 0.5in

{\bf Rafael Hern\'andez$^1$},\phantom{x} {\bf Konstadinos
Sfetsos}$^2$\phantom{x} and\phantom{x} {\bf Dimitrios Zoakos}$^2$
\vskip 0.1in

${}^1\!$
Theory Division, CERN\\
CH-1211 Geneva 23, Switzerland\\
{\footnotesize{\tt rafael.hernandez@cern.ch}}

\vskip .1in

${}^2\!$
Department of Engineering Sciences, University of Patras\\
26110 Patras, Greece\\
{\footnotesize{\tt sfetsos@des.upatras.gr, dzoakos@upatras.gr}}\\

\end{center}

\vskip .4in

\centerline{\bf Abstract}

\no
We summarize our recent work on supergravity backgrounds dual to part of the Coulomb
branch of ${\cal N}=1$ theories constructed as marginal deformations
of ${\cal N}=4$ Yang--Mills. In particular, we present a summary of the behaviour of the heavy
quark-antiquark potential which shows confining behaviour in the IR as well as of
the spectrum of the wave equation. The reduced supersymmetry
is due to the implementation of T-duality in the construction of the deformed supergravity
solutions.
As a new result we analyze and explicitly solve the Killing spinor equations of the
$\cN=1$ background in the superconformal limit.

\vfill

\noindent
\small{Proceedings of the RTN Workshop {\em{``Constituents, Fundamental Forces
and Symmetries of the Universe''}},
Corfu Summer Institute, 20-26 September 2005, to appear in Fortsch. Phys.
Based on a talk given by K.S.}

\end{titlepage}
\vfill
\eject

\def\baselinestretch{1.2}
\baselineskip 20 pt
\noindent

\def\tT{{\tilde T}}
\def\tg{{\tilde g}}
\def\tL{{\tilde L}}


\section{Introduction}

${\cal N}=4$ supersymmetric Yang--Mills admits a three parameter
family of exactly marginal deformations with a global $U(1)^3$
symmetry preserving ${\cal N}=1$ supersymmetry \cite{LS}. The
gravity dual of the ${\cal N}=4$ theory deformed by operators of
the form $\hbox{Tr } [\Phi_1 \{\Phi_2,\Phi_3\}]$ can be
constructed using an $SL(2,\mathbb{R}) \in SL(3,\mathbb{R})$
symmetry of the complete $SL(3,\mathbb{R}) \times SL(2,\mathbb{R})$ duality
group of type-IIB supergravity compactified on a two-torus \cite{LM} (see also
\cite{NiaPre}-\cite{Berenstein}). When the deformation parameter is real,
the marginally deformed background can be obtained
through a sequence of T-duality transformations and coordinate
shifts. As this derivation applies to supergravity backgrounds
preserving at least a global symmetry group isomorphic to
$U(1)^3$, we can extend it to find the marginal deformations of
solutions away from the conformal point. The Coulomb branch of the
${\cal N}=4$ theory arises when the $SO(6)$ scalar fields acquire
non-vanishing expectation values. These Higgs expectation values
correspond on the gravity side to multicenter distributions of
branes. In \cite{HSZ} we employed a sequence of T-dualities and
coordinates shifts on multicenter
solutions to find gravity duals for the Coulomb branches of
marginally deformed ${\cal N}=4$ Yang--Mills. In this note we will
focus on the deformation of uniform continuous distributions of
D3-branes on a disc and on a three-dimensional spherical shell
preserving a global $SO(4) \times SO(2)$ symmetry group. We will first
briefly present the marginally deformed backgrounds. We will then
investigate the supersymmetry preserved by these deformed solutions by
explicitly constructing the Killing spinor. We also
include a summary of the evaluation of the Wilson loop operator
along the deformed background, and an analysis of the spectra of
massless excitations after solving the corresponding Laplace
equation. The reader should consult \cite{HSZ} for further
details and related references.


\section{Marginally deformed backgrounds}

The supergravity solutions describing the Coulomb branch of ${\cal N}=4$ supersymmetric
Yang--Mills at strong 't~Hooft coupling involve a metric and a self-dual 5-form as the only
non-trivial fields. The metric has the form
\ba
ds_{10}^2 = H^{-1/2} \eta_{\m\n}dx^\m dx^\n + H^{1/2} dx_i dx_i \ ,
\qq \m=0,1,2,3\ ,\quad
i=1,2,\dots , 6\ ,
\label{mutll}
\ea
the self-dual 5-form is given by
\be
F_5 = dA_4+*_{10} dA_4\ ,\qq (dA_4)_{0123i}=-\del_i H^{-1} \
\label{lab1}
\ee
and the dilaton is a constant $\Phi_0$. The solution is completely characterized by a
harmonic function in $\mathbb{R}^6$. We are interested in the field theory limit
in which
the solution asymptotically becomes $AdS_5\times S^5$ with each factor having radius
$R=(4\pi g_sN)^{1/4}$, in string units.

\no
In order to obtain the deformed background we first split the metric
into a seven-dimensional piece, and a three-torus parametrized by some angles
$\phi_i$ (with $\phi_i\in (0,2 \pi)$),
\be
ds^2_{10} = G_{IJ}(x)dx^I dx^J + \sum_{i=1}^3 z_i(x) d\phi_i^2\ ,
\qq I=1,2,\dots , 7 \ .
\label{klw}
\ee
The seven-dimensional metric will not depend on the three angles.
The self-dual 5-form is then given by
\be
F_5 = d C^{(1)} \wedge d\phi_1 \wedge d\phi_2\wedge d\phi_3
+ {1\ov \sqrt{z_1z_2z_3}} *_7 d C^{(1)} \ ,
\label{lab2}
\ee
for some 1-form $C^{(1)} = C_I^{(1)} dx^{I}$.
Compatibility of \eqn{lab1} with \eqn{lab2} shows that there must be a 4-form
$C^{(4)}$ such that $dC^{(4)}= *_7 dC^{(1)}/\sqrt{z_1z_2z_3}$. We
now change variables through~\cite{LM}
\be
\phi_1=\varphi_3-\varphi_2\ , \qq \phi_2=\varphi_1+\varphi_2+\varphi_3\ ,\qq
\phi_3=\varphi_3-\varphi_1\ ,
\ee
and perform a T-duality transformation along the $\varphi_1$ direction, a coordinate shift
$\varphi_2 \rightarrow \varphi_2 + \gamma \varphi_1$, and again a T-duality along the
$\varphi_1$ direction. The resulting background was found in \cite{HSZ}.
For our purposes here we write the metric in the following form
\ba
&& ds_{10}^2  = G_{IJ}dx^Idx^J + G (z_2+z_3) \Big[ d\varphi_1
+ \frac {z_2}{z_2+z_3} d \varphi_2 + \frac {z_2-z_3}{z_2+z_3} d\varphi_3 \Big]^2
\label{metric}\\
&& + \, G \big( z_1 + \frac {z_2 z_3}{z_2 + z_3} \big) \Big[ d\varphi_2
- \frac {z_1(z_2+z_3) -2z_2z_3} {z_1z_2+z_2z_3 + z_3z_1} d\varphi_3 \Big]^2
+ 9 \frac {z_1z_2z_3}{z_1z_2+z_2z_3+z_3z_1} d\varphi_3^2 \ .
\nonumber
\ea
We will use a frame basis $e^a$, where
the indexes running from $0,1,\dots ,6$ correspond to some frame basis for the
seven-dimensional metric $G_{IJ}$, whereas $7,8$ and $9$ correspond to the natural frame
read off from \eqn{metric}, i.e. $e^{i+6}=(\dots)[d\varphi_i+\dots]$.
For the various forms and the dilaton supporting the solution we have
\ba
B \! & = & \! \gamma (z_1z_2 + z_2z_3 + z_3z_1)^{1/2} \, e^7\wedge e^8 \ ,
\nonumber \\
A^{(2)} \! & = &  3 \g d\varphi_3\wedge C^{(1)}\ ,
\nonumber\\
F_5 \! & = & \! 3 G d C^{(1)} \wedge d\varphi_1 \wedge d\varphi_2\wedge d\varphi_3
+ {1\ov \sqrt{z_1z_2z_3}} *_7 dC^{(1)} \ ,
\label{fosu}
\\
e^{2\Phi} \! & = & \! e^{2\Phi_0} G\ .
\nonumber
\ea
For notational convenience we have defined $G^{-1}=1+\g^2 (z_1z_2+z_1z_3+z_2z_3)$.
We also note that the
supergravity description remains valid at a generic point of the manifold if
\ba
R\gg 1\qq {\rm and}\qq   \g R^2 \equiv \hat \g \ll R\ .
\label{kfgsu}
\ea
The latter condition is also sufficient for the 2-torus parametrized by $\varphi_{1,2}$
to remain much larger than the string scale after the T-dualities. Finally, note that the
periodicities of the angular variables $\phi_i$ remain intact in the deformed background.

\subsection{The $SO(4)\times SO(2)$ background}

We will now present the marginally deformed $SO(4) \times SO(2)$ background in the case where the
D3-branes are uniformly distributed on a disc of radius $r_0$. The original metric is of the form
\eqn{mutll} with the flat metric in $\IR^6$ given by
\ba
ds^2_{\IR^6} & = &{r^2+r_0^2\cos^2\th\ov r^2+r_0^2}\ dr^2
+ (r^2+r_0^2\cos^2\th) d\th^2
+ (r^2+r_0^2) \sin^2\th d\phi_1^2
\nonumber\\
&& +\ r^2 \cos^2\th
(d\psi^2 + \sin^2\psi\ d\phi_2^2 +  \cos^2\psi\ d\phi_3^2) \ ,
\label{rhef}
\ea
where the 3-sphere line element is
\ba
d\Om_3^2 = d\psi^2 + \sin^2\psi\ d\phi_2^2 +  \cos^2\psi\ d\phi_3^2
\label{lpj2}
\ea
and the harmonic function reduces in this case to
\ba
H =  {R^4\ov r^2 (r^2+r_0^2 \cos^2\th)}\ .
\label{dj32}
\ea
The forms necessary to compute the NS-NS and R-R field strengths are
\ba
C^{(1)}& = & {R^4} {r^2+r_0^2\ov r^2+r_0^2 \cos^2\th}\ \cos^4\th\sin\psi
\cos\psi \ d\psi\  ,
\nonumber\\
C^{(4)}& =& -R^{-4}r^2 (r^2+r_0^2\cos^2\th)\
dt \wedge dx_1\wedge dx_2\wedge dx_3 \ .
\ea
The above allow to read off, with the aid of \eqn{klw}, the coordinates $z_1$, $z_2$ and $z_3$,
and from them to compute
\be
G^{-1}=1 + \hat \g^2 {\cos^2 {\th} \ov r^2+r_0^2 \cos^2 {\th}}\left[
(r^2+r_0^2) \sin^2 {\th} + r^2 \cos^2 {\th} \sin^2 {\psi} \cos^2 {\psi} \right] \ .
\ee
The case of a distribution of branes on the surface of a sphere of radius $r_0$ follows from
analogous expressions, after we replace $r_0^2 \rightarrow - r_0^2$.


\section{Supersymmetry}

We will now analyze in detail the supersymmetry of the marginally deformed backgrounds,
and describe the origin of the reduced supersymmetry of the solutions.
We will need the type-IIB
supergravity Killing spinor equations corresponding to the gravitino and dilatino
variations. In the string frame they read (see, for instance, \cite{Bergshoeff})
\ba
&& D_\mu \e - \frac{1}{8}\not\!\! H_{\mu}\sigma^3 \e +
\frac{1}{16}e^{\Phi} \sum_{n=1}^5 \frac{1}{(2n-1)!}
\not\!G^{(2n-1)}
\Gamma_\mu P_{n}\e   = 0 \ ,
\nonumber \\
&& \not\! \partial \Phi \e - \frac{1}{12}
\not\!\! H \sigma^{3} \e + \frac{1}{4}e^{\Phi} \sum_{n=1}^5 \frac{(n-3)}{(2n-1)!}
\not\! G^{(2n-1)} P_{n} \e  =  0\  ,
\ea
where $P_n=\s_1$ for even $n$ or $P_n=i\s_2$ when $n$ is odd, with $\s_i$ the Pauli matrices,
and we have defined $G^{(2n+1)}=dA^{(2n)}-H \wedge A^{(2n-2)}$.

\subsection{The supersymmetry breaking mechanism and T-duality}

We first present the mechanism responsible for the supersymmetry breaking from
$\cN=4$ to $\cN=1$ when deforming the original background. We will concentrate on
the background with $SO(4)\times SO(2)$ global symmetry in the disc case, and summarize
the relevant results in \cite{HSZ}. The solution for the Killing spinor in the
undeformed case can split into a part which is a singlet of the $U(1)$ rotations
corresponding to the angles $\varphi_1$ and $\varphi_2$, and a part orthogonal to that.
After the T-dualities and the coordinate shift only this part survives and remains a Killing spinor of
the deformed theory. For any multicenter metric of the form \eqn{mutll} the Killing spinor is
\be
\e = H^{-1/8} \e_0\ ,
\ee
with $\e_0$ a spinor subject to the projection
\ba
i \G^{0123}\e_0=\e_0\ ,
\label{hfrp}
\ea
where the indexes refer to the directions along the brane. This projection will not be necessary
in the conformal case in which $r_0=0$. The spinor $\e_0$ is determined to be
\be
\e_0 = e^{\ha f(r,\th)\G_{12}}
e^{{\psi\ov 2}
\G_{13}} e^{\ha \phi_i\s_i}
\bar \e_0\ ,
\ee
where $\bar \e_0$ is a constant spinor and where we have defined
\be
f(r,\th)=\tan\inv \left({r \tan {\th} \ov (r^2+r_0^2)^{1/2}}\right)\  \:\:
\hbox{and}\ \:\: \s_1=\G_{24}\ ,\quad \s_2=\G_{35}\ ,\quad \s_3= \G_{16}\ .
\label{skpi}
\ee
The calculations leading to the above expressions have been performed in a frame
which is read off directly from \eqn{rhef} (including the harmonic function),
that is $e^1=(\dots) dr$, $e^2= (\dots) d\th$, $e^3=(\dots) d\psi$ and $e^{i+3}=(\dots)d\phi^i$.
After rewriting the spinor $\e_0$ in the $\varphi_i$ coordinate system and a simple computation
we find that the required spinor invariant under variations of $\varphi_{1}$ and $\varphi_{2}$ is
given by
\ba
\e_{0,{\rm inv}} =  e^{\ha f(r,\th)  \G_{12}} e^{{\psi\ov 2} \G_{13}}
e^{{3\ov 2} \s_3\varphi_3} \bar \e_{0,{\rm inv}}\ .
\label{skpi8}
\ea
The constant spinor $\bar \e_{0,{\rm inv}}$ in terms of $\bar \e_0$ is given by
\be
\bar \e_{0,{\rm inv}}={1\ov 4} (\II-\s_1\s_2-\s_1\s_3-\s_2\s_3)\bar \e_0\ ,
\ee
where the prefactor acts as a projector and by construction we have
\ba
\s_1\bar \e_{0,{\rm inv}}=\s_2 \bar \e_{0,{\rm inv}}= \s_3 \bar\e_{0,{\rm inv}}\ .
\label{projj}
\ea

\subsection{The explicit Killing spinor in the conformal case}

We will now construct the ten-dimensional Killing spinor in the deformed background.
This computation is quite involved technically, so that for
simplicity we will only consider in some detail the conformal limit, $r_0=0$.
The non-vanishing components of the spin connection for \eqn{metric} are
\ba
\om^{i4} & = & e^i\ , \quad i=1,2,3\ , \quad \om^{56} =
{\textrm{s}_{\th} \ov \textrm{c}_{\th}} \, e^6 \ , \nonumber
\\
\om^{57} & = & G{\textrm{s}_{\th} \ov \textrm{c}_{\th}}
\left[1+\hat\gamma^2\textrm{c}_{\th}^4\left(1-\textrm{s}_{\psi}^2\textrm{c}_{\psi}^2\right)\right]e^7 \ , \nonumber
\\
\om^{58} & = & -{G \textrm{s}_{\th}\textrm{c}_{\th}^3 \ov u^2}
\left[1-\textrm{s}_{\psi}^2\textrm{c}_{\psi}^2 + \hat\gamma^2  {u^2 \ov \textrm{c}_{\th}^2}
\right] e^8 + {\sqrt{G} \textrm{s}_{\psi} \textrm{c}_{\psi} \textrm{c}_{\th}^2 \ov u^2}\,e^{9} \ , \nonumber
\\
\om^{59} & = & {\sqrt{G} \textrm{s}_{\psi} \textrm{c}_{\psi} \textrm{c}_{\th}^2 \ov u^2} \, e^{8} +
{\textrm{c}_{\th} \ov \textrm{s}_{\th}}{\textrm{}s_{\th}^4-\textrm{c}_{\th}^4 \textrm{s}_{\psi}^2 \textrm{c}_{\psi}^2 \ov
u^2} \, e^{9} \ , \nonumber
\\
\om^{67} & = & {1 \ov 4} \, \hat\gamma^2G \textrm{c}_{\th}^3 \textrm{s}_{4\psi}e^7
-{\textrm{s}_{\psi} \textrm{c}_{\psi} \textrm{c}_{\th} \ov u }\,e^8
+{\sqrt{G} \textrm{s}_{\th} \ov u} \, e^{9} \ ,\nonumber
\\
\om^{68} & = & -{ \textrm{s}_{\psi} \textrm{c}_{\psi} \textrm{c}_{\th} \ov u} \, e^7 -
{G \textrm{c}_{\th}^3 \textrm{s}_{4\psi} \ov 4u^2} \, e^8
- {\sqrt{G} \textrm{c}_{2\psi} \textrm{s}_{\th} \textrm{c}_{\th}^2 \ov u^2} \, e^{9} \ ,
\\
\om^{69} & = & - {\sqrt{G} \textrm{s}_{\th} \ov u} \, e^{7} -
{\sqrt{G} \textrm{c}_{2\psi} \textrm{s}_{\th} \textrm{c}_{\th}^2 \ov
u^2} \, e^{8} - {2 \textrm{s}_{\th}^2 \textrm{c}_{\th}\textrm{c}_{2\psi}
\ov \textrm{s}_{2\psi}}\,{1 \ov u^2} \, e^{9} \ , \nonumber
\\
\om^{78} & = & -{ \textrm{s}_{\psi} \textrm{c}_{\psi} \textrm{c}_{\th} \ov u} \ e^6\ , \quad
\om^{79} = - {\sqrt{G} \textrm{s}_{\th} \ov u} \, e^{6} \ , \nonumber
\\
\om^{89} & = & {\sqrt{G} \textrm{s}_{\psi} \textrm{c}_{\th}^2 \ov
u^2} \, e^{5} - {\sqrt{G} \textrm{c}_{2\psi} \textrm{s}_{\th} \textrm{c}_{\th}^2 \ov u^2} \ e^{6} \ ,
\nonumber
\ea
where we have introduced the notation ${\rm c}_{\a} \equiv \cos \a$ and
${\rm s}_{\a} \equiv \sin \a$ and
\be
u \equiv {\rm c}_{\th} \sqrt{{\rm s}_{\th}^2+{\rm c}_{\th}^2 {\rm s}_{\psi}^2{\rm c}_{\psi}^2}\ .
\ee
From the gravitino variation along the brane directions and $\m=r$ we get
the following expression for the spinor \cite{spinor}
\ba
\e=e^{-{1 \ov 2}\ln{r}\G^{(5)}A}
\left[\II - {1 \ov 2}x^{\a}\G_{\a5}\left(\II +  \G^{(5)} A \right)\right]\eta \ ,
\ea
where $\eta$ is a spinor that could depend on everything but the worldvolume coordinates
$x^{\a}$ and $r$. Also $\G^{(5)}=i\G^{0123}$ and $A$ is defined as
\be
A \equiv \sqrt{G} \left (\II - \hat\gamma u \G^{78} \ast \right)
= e^{-\tan^{-1} ( \hat\g u) \G^{78}\ast } \ ,
\ee
with $*$ the complex conjugation operator. The gravitino variations along
$\m=\th$, $\m=\psi$ and $\m=9$ determine
\be
\eta = e^{{1 \ov 2}\tan^{-1}{\hat\g u}\G^{78}\ast}e^{-\frac {1}{2} \tan^{-1} \left(
2\tan\th/\sin 2\psi \right) \G_{89} }
e^{-\frac {1}{2} \th \G_{45}\G^{(5)}}
e^{\frac {1}{2} \psi (\G_{78} - \G_{46}\G^{(5)}) } e^{{3\ov 2} \G_{68}\varphi_3}\eta_0 \ ,
\label{spopi}
\ee
where $\eta_0$ is a constant spinor satisfying two projections arising from the
supersymmetry variation for the dilatino,
\be
\G_{47}\G^{(5)} \eta_0 = \G_{68} \eta_0 = \G_{59} \eta_0 \ .
\ee
In addition, from the gravitino variations along $\mu=7$ and $\mu=8$ we conclude that
there is no dependence of the spinor on $\varphi_1$ and $\varphi_2$. Note that the spinor
\eqn{spopi} does not reduce precisely to that in \eqn{skpi8} as $\hat\g\to 0$, since the
corresponding frames are different.


\section{Wilson loops}

The marginally deformed backgrounds in the Coulomb branch contain a very rich structure.
In \cite{HSZ} we probed the geometry of the deformation by evaluating the expectation
value of the Wilson loop operator along the transversal space to the worldvolume of the
branes. The Wilson loop can be computed by minimizing the Nambu--Goto action for a
fundamental string in a given supergravity background \cite{loop}. We will in particular
take the string to stretch along a trajectory given by
\be
\th=0\ ,\qq \psi={\pi\ov 4} \ ,
\qq \phi_2=\phi_3\equiv \phi\  , \qq x_{2,3}=\hbox{constant} \ ,
\ee
which is consistent with the equations of motion if the conserved angular momenta
coincide, $l_{\phi_2}=l_{\phi_3} \equiv l$. In the absence of a Higgs expectation value,
$r_0 \rightarrow 0$, we recover the expected Coulombic behaviour for the heavy
quark-antiquark potential. In the non-conformal case we will only concentrate on the case
of the disc distribution. Then the behaviour of the quark-antiquark potential
depends on the relation between the various parameters of the theory. At a finite value
of the length of the loop,
$L_{\hbox{\footnotesize{fin}}} = \frac {\pi R^2}{r_0} \sqrt{1-l^2}$, the
potential goes to zero,
\ba
E_{q\bar q} \! & \simeq & \! - {1-(1+\hat \g^2/4)l^2\ov 4(1-l^2)} \ r_0^3
\left(L_{\rm fin}-L\ov \pi R^2\right)^2 \ .
\ea
If $(1+\hat \g^2/4)l^2<1$ the potential vanishes monotonically. When $(1+\hat \g^2/4)l^2>1$
the potential becomes positive, reaches a maximum and then becomes zero again. However,
beyond the maximum the force between the quark and the antiquark becomes repulsive.
To avoid this unphysical region we take the limit of a large deformation
parameter, $\hat \g\gg 1$, and simultaneously take the angular parameter $l\to 1$.
This introduces a hierarchy of widely separated scales
and in particular we note that
\ba
{r_0\ov \hat \g} \ll r_0 \ll {r_0\ov \sqrt{1-l^2}}\ .
\ea
The above limit avoids the unphysical region by removing the piece in the deep IR. The
conformal region is also removed by avoiding probing with energies in the deep UV, that
is extremely higher than the Higgs expectation value. In this way, for small values of the
separation distance $\bar{L}=\frac {L}{R^2\sqrt{1-l^2}}$, we find a linear behaviour
\be
E_{q\bar q} \simeq {r_0^2 \ov 2 \pi} \bar L \ ,\quad {\rm as}
\quad {\sqrt{1-l^2}\ov r_0}\ll \bar L \ll {1\ov r_0} \
\label{liin}
\ee
and a logarithmic dependence for large values of $\bar{L}$,
\be
E_{q\bar q} \simeq {r_0 \ov \pi} \ln \left(r_0\bar L\right)
\ , \quad {\rm as}\quad {1\ov r_0}\ll \bar L \ll {\hat \g\ov r_0}\ .
\label{logr}
\ee
A logarithmic form for a confining potential instead of a linear
one was suggested long time ago in order to explain quarkonium spectra with energy level spacing
independent of the particle mass \cite{quigg}. The various different behaviours are depicted in
Figure 1.

\begin{figure}[h!]
\begin{center}
\includegraphics[height= 5 cm,angle= 0]{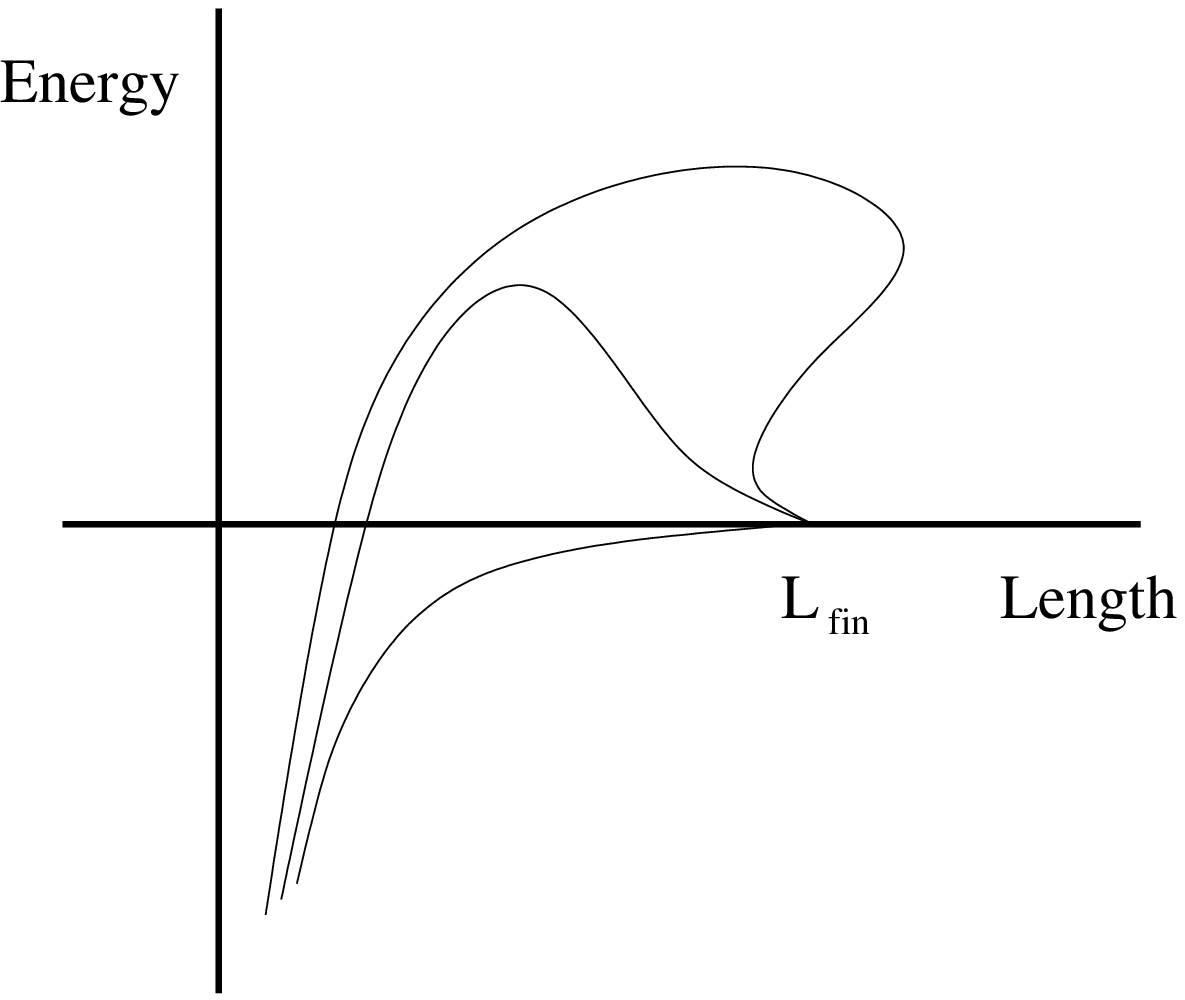}\qq\qq
\includegraphics[height= 5 cm,angle= 0]{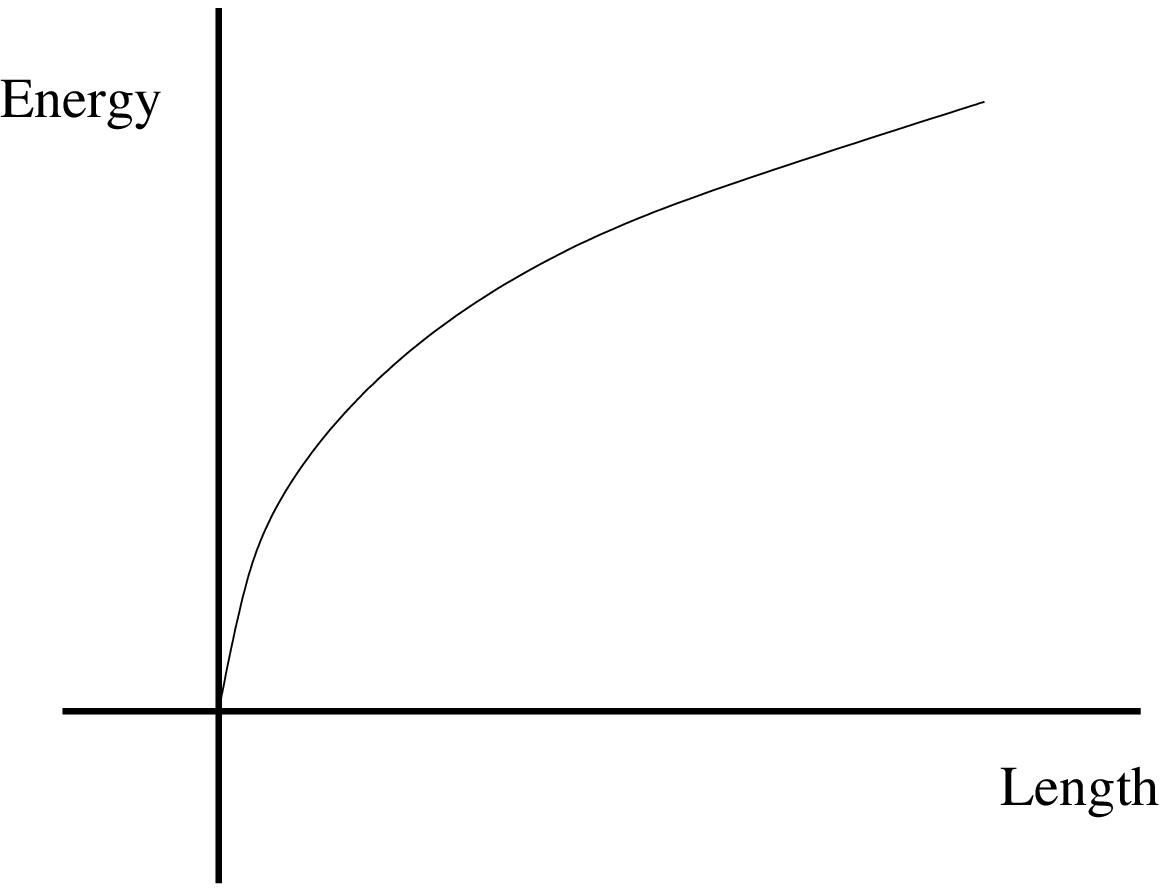}
\end{center}
\caption{The energy as a function of the quark-antiquark separation distance $L$. 
\underline{Left}: As the value of $\hat{\gamma}$ increases,
the shape of $E_{q\bar{q}}$ is modified
from the lower to the upper curve.
The result is trusted until the maximum energy is reached.
\underline{Right}: In the limit $\hat \g\to \infty$, $l\to 1$
and for large separations we find the logarithmic confining dependence \eqn{logr}.
For smaller separations the behaviour is linear, \eqn{liin}.
}
\label{fig1}
\end{figure}


\section{The wave equation}

The study of the massless field equations in a supergravity background is a problem
of great interest because the spectrum of fluctuations after the AdS/CFT correspondence
corresponds to gauge theory operators. In \cite{HSZ} we found that the marginal deformation
in the massless scalar field equation arises as an additional term for the corresponding
undeformed equation,
\be
\Box \Psi = \Box_{\g=0} \Psi +\g^2 \l_{ij}\ \del_i \del_j \Psi = 0 \  \ ,
\label{Laplace}
\ee
where we have introduced the symmetric matrix
\be
\l_{ij}\equiv  {z_1z_2 +z_1z_3 +z_2 z_3\ov z_i}\d_{ij}
-{z_1 z_2 z_3 \ov z_i z_j} \ .
\ee

\subsection{The undeformed case}

It turns out that in the absence of deformation, the Laplace equation
provides a set of differential equations
that can be solved by transforming them into Schr\"odinger equations whose potential is
determined by the geometry of the supergravity background. We first
separate variables on the Laplace operator through the plane-wave ansatz
\ba
\Psi = {1\ov (2\pi)^{5/2}} e^{ik\cdot x} e^{i n\phi_1}
\Psi_{S^3}(\psi,\phi_2,\phi_3) \psi^{(1)}(\th) \psi^{(2)}(r) \ , \qq n\in {\mathbb Z} \ .
\label{djwh}
\ea
The equation for $\Psi_{S^3}$ is the usual eigenvalue equation on $S^3$ with eigenvalue
$-l(l+2)$, $l=0,1,\dots$.
The one for $\psi^{(1)}(\th)$ becomes the Jacobi differential equation provided that
a separation variable $E$ is quantized as
\ba
E_{m,l,n}=(l+|n|+2m)(l+|n|+2 m +4)\ ,\qq m= 0,1,\dots\ .
\label{jhgf}
\ea
This parameter enters into
the radial equation for $\psi^{(2)}(r)$ which is the only one sensitive to the details of
the geometry. We must therefore consider three possible cases. \\
\no
{\bf The conformal limit:} Once we set $r_0=0$ the differential equation
for $\psi^{(2)}(r)$ can be
transformed into a Schr\"odinger equation with potential
\be
V(z)= {15/4+E_{m,l,n}\ov  z^2}
\ee
and eigenvalue $M^2R^4$, where $z = {1/r}$.
As this
is a positive definite potential, which vanishes for large values of $z$, we find a continuous
spectrum with no mass gap. \\
\no
{\bf The disc:} When the D3-branes distribute uniformly
on a disc of radius $r_0$ we transform the problem into a Schr\"odinger
equation with potential
\be
V(z)= (l+1)^2-{n^2-1/4\ov \cosh^2 z} + {15/4+E_{m,l,n}\ov \sinh^2 z} \ ,
\ee
with $\sinh z = r_0/r$ and eigenvalue $M^2R^4/r_0^2$. This potential decreases monotonically
from arbitrarily large positive values to the constant $(l+1)^2$, as $z$ varies from $0$ to $\infty$,
and belongs to the family of P\"oschl--Teller potentials in quantum mechanics of type II.
We therefore find a continuous spectrum with mass gap given by
\be
M_{{\rm gap},l}= (l+1) {r_0\ov R^2} \ .
\ee
It turns out that this gap is $(l+1)^2$ degenerate.

\no
{\bf The sphere:} In this case, after changing to a new radial variable $\sin z = r_0/r$,
we transform the problem into a Schr\"odinger
equation with potential
\be
V(z) = -(l+1)^2+{n^2-1/4\ov \cos^2 z} + {15/4+E_{m,l,n}\ov \sin^2 z} \ ,
\ee
and eigenvalue $M^2R^4/r_0^2$, which belongs to the family of P\"oschl--Teller
potentials of type I. The mass eigenvalue turns out to be quantized as
\ba
M^2_{k,m,l,n}= 4(k+m+|n|+1)(k+m+|n|+l+2){r_0^2\ov R^4}\ ,
\label{mmeig}
\ea
with degeneracy $(l+1)^2(k+m+|n|+1)^2$.

\subsection{The deformed case}

The deformed backgrounds include the second term on the right hand side of (\ref{Laplace}).
As this term breaks the $SO(4)$ spherical symmetry unless we focus on solutions independent of
the angles $\phi_2$ and $\phi_3$, we must now consider the ansatz (\ref{djwh}) but
with $\Psi_{S^3}$ having no dependence on $\phi_2$, $\phi_3$ (a slight generalization can be
found in \cite{HSZ}). It turns out that then the quantum number $l$ has
to be an even integer, so that in the rest we replace it by $2 l$.
The radial equation for $\psi^{(2)}(r)$ is basically the same
as in the undeformed case. It is the equation for $\psi^{(1)}(\th)$ that receives
the major effect of the deformation. As before the equation can be transformed into
a Schr\"odinger equation with potential
\be
V(\th)= -4+{n^2-1/4\ov \sin^2\th}   + {4l(l+1) +3/4\ov \cos^2 \th}\
+ n^2 \hat \g^2 \cos^2\th \ .
\label{VIno}
\ee
When $n=0$ we recover the undeformed limit. Solving the equation for $n \neq 0$
is a difficult
task, because there is a change in the nature of the singularity at infinity
of the corresponding
differential equation which is of the Fuchsian type.
Perturbation theory or an asymptotic expansion can still be used,
for small or large values of $n^2 \hat\g^2$, respectively. For arbitrary values of the
deformation parameter we may appropriately
redefine $\psi^{(1)}$ (for details we refer the reader to \cite{HSZ})
to reduce the problem to a confluent form of the Heun differential equation
(see, for instance, \cite{Ronveaux}). The Heun differential equation is known to be related
to the BC$_1$ Inozemtsev system, which is a one-particle quantum mechanical model in one dimension
with an elliptic potential \cite{Inozemtsev}.
In fact, after a trigonometric limit the Inozemtsev
system becomes a Schr\"odinger problem with a trigonometric P\"oschl--Teller potential, which is
nothing but (\ref{VIno}) with $\hat \g=0$. The complete potential can still be recovered through a
generalized form of the trigonometric limit \cite{oshima,HSZ}. The importance of the relation
to the Inozemtsev system is that this is an integrable model. Therefore, the Bethe ansatz method
can be used to find solutions to the deformed differential equation. We expect that
further progress
can be made along this line.


\section{Conclusions}

The exactly marginal deformations of ${\cal N}=4$ supersymmetric Yang--Mills by operators of
the form $\hbox{Tr } [\Phi_1 \{\Phi_2,\Phi_3\}]$ amount on the gravity side of the AdS/CFT
correspondence to a deformation on the transversal space to the worldvolume of the branes. We
have explored the geometry of this deformation both in the conformal limit and for a continuous
distribution of D3-branes with $SO(4) \times SO(2)$ global symmetry group. We have in particular
presented the mechanism responsible for the breaking of supersymmetry from ${\cal N}=4$ to
${\cal N}=1$ as the supergravity background gets deformed, and constructed the ten-dimensional
Killing spinor. By evaluating the Wilson loop operator we have found regimes where the
quark-antiquark interaction is completely screened, or where a confining behaviour arises.
Finally,
we have also performed a detailed analysis of the spectra of massless excitations in the
deformed $SO(4) \times SO(2)$ background. We have found solutions to the Laplace equation
by transforming the corresponding differential equations into Schr\"odinger problems.
The rich structure and underlying geometry of the deformations present themselves as a
promising path for further investigations.


\vskip .4in

\centerline{ \bf Acknowledgments}

\no
K.S. and D.Z. would like to thank the organizers of the RTN Workshop in Corfu, Greece,
20-26 September 2005, where parts of this research were presented, for hospitality and
for creating a productive scientific atmosphere.
K.S. and D.Z. acknowledge the financial support provided through the European Community's
program ``Constituents, Fundamental Forces and Symmetries of
the Universe'' with contract MRTN-CT-2004-005104,
the INTAS contract 03-51-6346 ``Strings, branes and higher-spin gauge fields'', as well as
the Greek Ministry of Education programs $\rm \P Y\Th A\G OPA\S$ with contract 89194 and
$\rm E\Pi A N$ with code-number B.545.
In addition, D.~Z. acknowledges the financial support
provided through the Research Committee of the University of Patras for a ``K.~Karatheodory''
fellowship under contract number 3022.


\end{document}